\documentclass[aps,pra,twocolumn,groupedaddress]{revtex4}
\usepackage{graphicx}
\usepackage{amsmath}
\usepackage{amsfonts}
\usepackage{bm}
\usepackage{color}
\newcommand{\ket}[1]{\ensuremath{\left|{#1}\right\rangle}}

\newcommand{\sign}{\ensuremath{{\mathrm{sgn}}}}

\newcommand{\beq}{\begin{equation}}
\newcommand{\eeq}{  \end{equation}}
\newcommand{\bea}{\begin{eqnarray}}
\newcommand{\eea}{  \end{eqnarray}}
\newcommand{\bit}{\begin{itemize}}
\newcommand{\eit}{  \end{itemize}}

\begin{document}


\title{Observation of tunable Popescu-Rohrlich correlations through post-selection of a gaussian state}

\author{D. S. Tasca}
\email[]{tasca@if.ufrj.br}
 \affiliation{Instituto de F\'{\i}sica,
Universidade Federal do Rio de Janeiro, Caixa Postal 68528, Rio de
Janeiro, RJ 21941-972, Brazil}
\author{S. P. Walborn}
\affiliation{Instituto de F\'{\i}sica, Universidade Federal do Rio
de Janeiro, Caixa Postal 68528, Rio de Janeiro, RJ 21941-972,
Brazil}
\author{F. Toscano}
 \affiliation{Instituto de F\'{\i}sica, Universidade Federal do Rio
            de Janeiro, Caixa Postal 68528, Rio de Janeiro, RJ 21941-972,
            Brazil}

\author{P. H. Souto Ribeiro}
\affiliation{Instituto de F\'{\i}sica,
Universidade Federal do Rio de Janeiro, Caixa Postal 68528, Rio de
Janeiro, RJ 21941-972, Brazil}
\date{\today}

\begin{abstract}
We show that non-local Popescu-Rohrlich correlations can be observed in the post-selected results of binned position measurements on a two-party gaussian state. Our experiment is based on the spatial correlations of entangled photons and lens systems. We obtain a maximum violation of the CHSH inequality of 3.42, which corresponds to the implementation of a non-local AND gate with success probability of 0.93. These results do not conflict with quantum mechanics due to the post-selection required, and open up the possibility of experimental investigation of fundamental aspects of Popescu-Rohrlich non-locality with a reliable and simple experimental setup.
\end{abstract}

\pacs{42.50.Xa,42.50.Dv,03.65.Ud}

\maketitle
In 1965, John Bell showed that quantum mechanics exhibits non-local correlations which can be stronger than any classical correlation \cite{bell65}.  Classical correlations obey the CHSH inequality  $S \leq 2$, where $S$ is known as the Bell parameter \cite{chuang00,chsh}.  For certain quantum  states, $S$ can exceed 2, while the maximum value obtainable is $2 \sqrt{2}$, and is known as Tsirelson's bound \cite{cirelson80}.  Popescu and Rohrlich (PR) have proposed and investigated correlations that could lead to the maximum possible value of  $S=4$ and therefore violate Tsirelson's bound. In their work, they have demonstrated that these PR correlations, produced by what are now known as PR boxes, are compatible with the causality imposed by special relativity, even though they are stronger
than any known classical or quantum correlation \cite{popescu94}.

\par
The PR box relates Alice and Bob's respective input bits $A$ and $B$ with their respective output bits $a$ and $b$ according to the relation $a \oplus b = A \cdot B$, where $\oplus$ stands for addition modulo 2.  This definition guarantees that $a=b$ whenever either $A$ or $B$ is zero, and that $a \neq b$ whenever $A=B=1$.  Also required for the definition of the PR box are that ({\it i}) Alice and Bob's marginal distributions are completely random: $P_A(a)=P_B(b)=1/2$ for all $A,B,a,b$; and ({\it ii}) satisfy relativistic causality, i.e $P_{A,B}(a)=P_{A,B^\prime}(a)$ for all $A,B,B^\prime, a$, etc\cite{popescu94}.  These basic restrictions guarantee that it is impossible to communicate directly via the PR box. 

\par
There has been much recent research focused around the identification of the properties of non-local correlations \cite{barrett05,brassard06,short06,masanes06,barrett07,vandam05}, and also in relating PR correlations to quantum entanglement \cite{cerf05,pitowsky08,brunner08},  and to quantum communication tasks \cite{scarani06}.  There has also been some effort in the investigation of the computational power of PR correlations in the context of non-local computation \cite{linden07,anders08}.
W. van Dam \cite{vandam05} and G. Brassard et al. \cite{brassard06} investigated the implications of the stronger-than-quantum non-locality in the field of communication complexity. W. van Dam showed that the correlations of the PR box allow for the solution of a two part distributed Boolean function with only one bit of classical communication between the parts. This implies that communication complexity would always be trivial if a physical system could present PR correlations. In [6] the authors
analyze a probabilistic version of the PR box. They concluded that a PR box operating with a success probability up to $90.8\%$,
implies in non-trivial communication complexity. However, contrary to what could be expected, this limit does not coincide
with Tsirelson's bound, since quantum mechanical correlations are equivalent to the implementation of a  probabilistic PR box
with maximmum sucess probability, or fidelity, of $85.4\%$. For classical correlations the maximal fidelity is $75\%$.

\par
Even though PR boxes produce non-physical correlations, in the sense that they cannot be produced by classical and quantum systems, probabilistic PR boxes can be simulated with high fidelities. Gisin has shown that local filtering of spin measurements can be used to surpass Tsirelson's bound, even in the case of classically correlated spin-1/2 states \cite{gisin96}. Marcovitch et al. have demonstrated that appropriately pre and post-selected ensembles of bipartite quantum states can exhibit stronger than quantum correlations  \cite{marcovitch07}. Chen et. al \cite{chen06} have surpassed the Tsirelson's bound by appropriately redefining measurements on three party GHZ states \cite{cabello02}.

\par
Here we report an experiment which demonstrates high fidelity PR correlations in variable post-selected dichotomic measurements, that are performed on a bipartite continuous-variable gaussian state. We surpass the non-trivial communication complexity limit, obtaining $93\%$ fidelity with respect to ideal PR correlations. We use twin pairs of photons produced in parametric down-conversion and perform measurements on the transverse spatial variables of the photons. This system is particularly interesting for the study of non-local correlations.  On the one hand, it presents genuine quantum correlations between the transverse variables of the photon pairs \cite{howell04,tasca08,tasca09}, demostrated through the violation of continuous variable separability and ERP-like correlation criteria \cite{duan00, mancini02,reid88}. On the other hand, because the quantum state of the photon pair can be   approximately described by a positive Wigner function, the non-local correlations  can only be observed through the measurement of observables with a particular Weyl-Wigner representation \cite{revzen05}. In the experiment reported here, we measure dichotomic spatial  observables that are not of this type.  However, we show that the PR correlations appear through a spatial filtering process. Moreover, we show that the strength  of the PR correlations  can be readily tuned. This manipulation of the correlations is a clear demonstration of how a detection loophole can be exploited to obtain a false violation in a Bell's inequality experiment.

Consider a pure bipartite state with gaussian Wigner function  $W(x_1,p_1,x_2,p_2)$, where $j=1,2$ and $x_j$ and $p_j$ refer to dimensionless position and momentum variables satisfying $[x_j,p_k]=i \hbar\delta_{jk}$.  Rotations in phase space can be realized through local operations given by
\begin{subequations}
\label{eq:rotations}
\begin{align}
\hat{x}_j\longrightarrow \hat{x}_j^\theta = \cos\theta \hat{x}_j + \sin \theta \hat{p}_j  \\
\hat{p}_j\longrightarrow \hat{p}_j^\theta = \cos\theta \hat{p}_j - \sin \theta \hat{x}_j,
\end{align}
\end{subequations}
and dichotomic measurements can be made through the sign operator $\sign(\hat{x}_j^{\theta})$.  The expectation value is \cite{gour04,revzen05}
\begin{equation}
\langle \sign(\hat{x}_1^{\alpha}) \sign(\hat{x}_2^{\beta})\rangle = \iint W(\xi_1^\alpha,\xi_2^\beta) \sign(x_1^{\alpha}) \sign(x_2^{\beta})d\xi_1^{\alpha} d\xi_2^{\beta},  
\label{eq:P}
\end{equation}  
where $\xi_j^\theta=(x_j^\theta,p_j^\theta)$.        
\par
The spatial degrees of freedom of photons produced by spontaneous parametric down conversion (SPDC) can be described to good approximation by a gaussian Wigner function \cite{chan07}.  In the simplest case, the quantum state of the photon pair can be written as
\begin{equation}
\ket{\psi}= A \iint  dq_1 dq_2 \exp{\left( -\frac{q_1^2+q_2^2}{ 2 \Delta^2} -\frac{q_1q_2}{ \gamma^2} \right )} \ket{q_1} \ket{q_2},  
\label{eq:state}
\end{equation}    
where $A$ is a normalization constant, $q_j=(s/\hbar)p_j$ (j=1,2) is the dimensionless transverse wave-vector of photon $j$ and $s$ is a constant with dimension of length \cite{tasca09}.  For $1/\gamma^2=0$ and nonzero $\Delta$ the state is separable, while for $\gamma^2=\Delta^2\longrightarrow0$ the state is a maximally entangled EPR state.    
Its Wigner function is given by a gaussian function and can be used according to Eq. (\ref{eq:P}). The rotations described in Eqs. (\ref{eq:rotations}) can be physically implemented for the transverse degrees of freedom of photons using optical lens systems designed to implement a fractional Fourier transform (FRFT)  \cite{ozaktas01,tasca08, tasca09} . In this case, each down converted  photon is emitted from the SPDC source, propagates through the FRFT system which implements a rotation with angles $\alpha (\beta)$, and is detected in a position $x_1(x_2)$ of the detection plane. 
\par
The probability of detecting photon 1 at position $x^{\alpha}_1$ and photon 2 at position $x^{\beta}_2$ is given by

\beq
R(x^{\alpha}_1,x^{\beta}_2)= \int \int W(\xi_1^\alpha,\xi_2^\beta)dp^{\alpha}_1 dp^{\beta}_2.
\eeq
We are interested in the conditional detection probabilities, for two special cases. The first is when $\alpha = \pi$, corresponding to an imaging system for photon 1:

\begin{align}
R_{\pi,\beta}(x_1,x_2)= & \left [\frac{\Delta^2\gamma^4}{\gamma^4-\Delta^4}\left(\frac{\cos^2\beta}{\Delta^2}+\Delta^2\sin^2\beta \right) \right]^{-1/2}  \times \nonumber \\
& \exp\left(\frac{-(\gamma^4-\Delta^4)x_2^2}{\Delta^2\gamma^4}\right ) \nonumber \\
& \exp \left(-\frac{(x_1-\frac{\Delta^2}{\gamma^2} \sin\beta x_2)^2}{\frac{1}{\Delta^2}\cos^2\beta + \Delta^2\sin^2\beta} \right),  \nonumber \\
\label{eq:p1}
\end{align}
where the subindexes ``$\pi$" and ``$\beta$" of the conditional probability $R$ refer to the phase-space rotations applied on the transverse coordinates of photons 1 and 2, respectively.

The second case is when $\alpha= \pi /2$, corresponding to a Fourier transform system for photon 1. In this case the detection probability is given by:

\begin{align}
R_{\frac{\pi}{2},\beta}(x_1,x_2)=& \left [{\Delta^4\gamma^8}{\sin^2\beta}+(\gamma^4-\Delta^4)^2\cos^2\beta  \right]^{-1/2}  \times \nonumber \\
&  \frac{(\gamma^4-\Delta^4)}{2\Delta}\exp\left(\frac{-(\gamma^4-\Delta^4)x_2^2}{\Delta^4\gamma^4}\right)  \times \nonumber \\
& \exp\left(\frac{\Delta^2\gamma^4(\gamma^4-\Delta^2)(x_1+\frac{\Delta^2}{\gamma^2} \cos\beta x_2)^2}{\Delta^4\gamma^8{\sin^2\beta}+(\gamma^4-\Delta^4)^2\cos^2\beta} \right).  \nonumber \\
\label{eq:p2}
\end{align}

Our experiment is designed to implement projections of the sign operators, after the phase space rotations through the FRFT systems. In practice this is done by collecting and detecting all photons falling on the upper ($+$) or lower ($-$) half of the detection plane and blocking the photons falling on the other half. To calculate the conditional probabilities that the signal photon is detected in the upper/lower half, conditioned to the detection of the idler in the upper/lower half, one should integrate the expressions given by Eqs. (\ref{eq:p1}) and (\ref{eq:p2}):

\begin{equation}
P_{\alpha\,\beta}(\pm,\pm)=(\pm1)(\pm1)\int\limits_{\pm r}^{\pm \infty}\int\limits_{\pm r}^{\pm \infty}R_{\alpha\,\beta}(x_1,x_2) dx_1 dx_2,
\end{equation}  where the parameter $r$ defines a {\it dark} region in the detection plane.

\begin{figure}
\includegraphics[width=6cm]{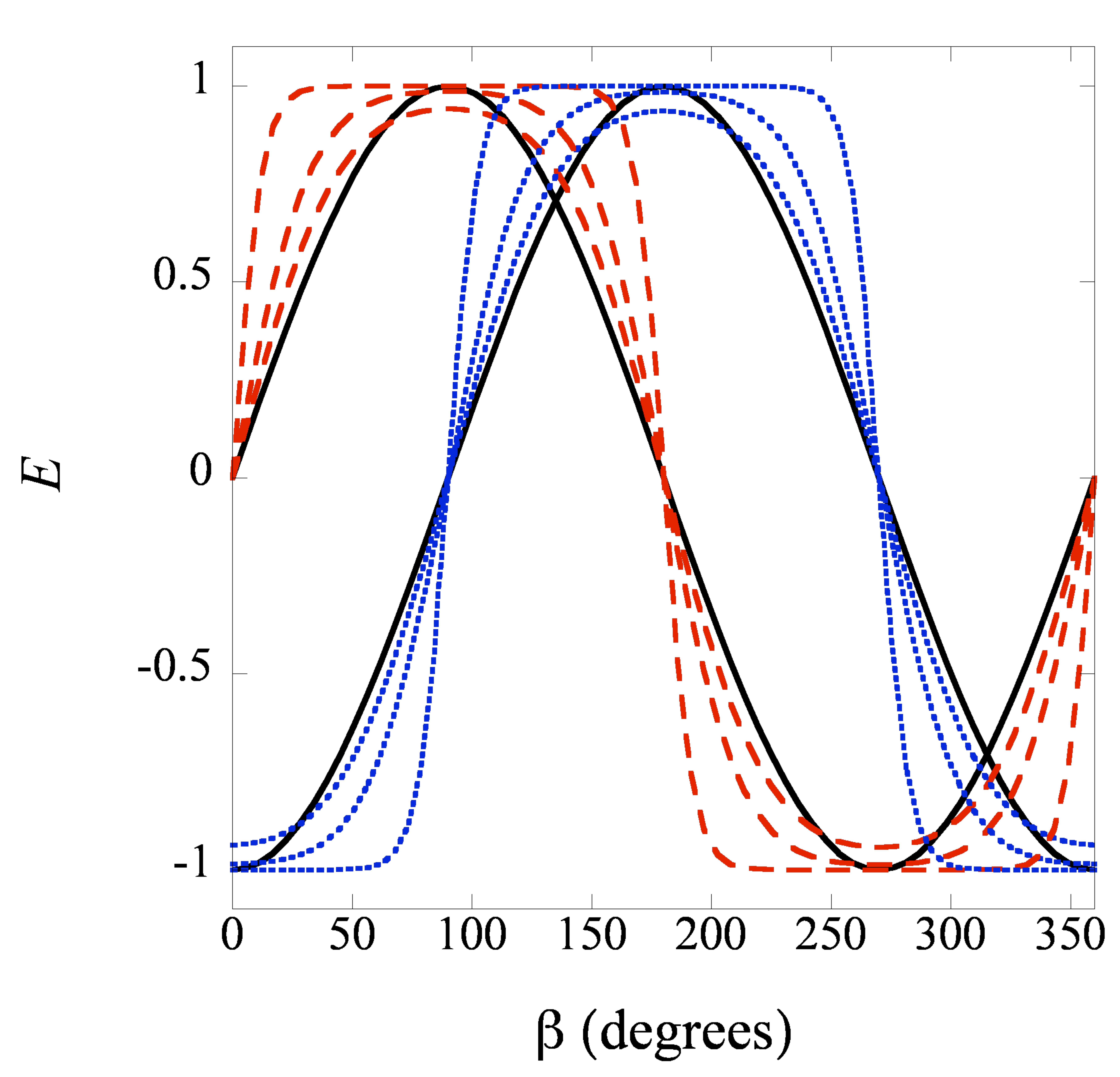}
 \caption{\label{fig:plots} (Color online) Correlation functions $E(\pi,\beta)$ (Red dashed curves)  and $E(\pi/2,\beta)$ (blue dotted curves)  as a function of $\beta$ for $r=0.75, r=1, r=2$. Also shown are the sine curves (black solid curves) corresponding to the maximum value obtained by quantum mechanics without post-selection.}
 \end{figure}

The variable dependent post-selection is achieved by choosing $r>0$, so that the photons which arrive in the region between $-r$ and $r$ are never detected.  The Bell parameter is defined as $\mathcal{S}=E(\alpha,\beta)+E(\alpha^\prime,\beta)+E(\alpha,\beta^\prime)-E(\alpha^\prime,\beta^\prime)$, where $E(\alpha,\beta)=P_{\alpha,\beta}(+,+)+P_{\alpha,\beta}(-,-)-P_{\alpha,\beta}(+,-)-P_{\alpha,\beta}(-,+)$. Here  ``$+$" \,and ``$-$" \, refer to the upper and lower regions of the detection plane respectively. Figure \ref{fig:plots} shows plots of $E(\pi,\beta)$ and $E(\frac{\pi}{2},\beta)$ as function of $\beta$ for $\Delta=5/4$ and $\gamma=3/4$ and several values of $r$.  For comparison, the black solid sine curve shows the results for a maximally entangled singlet state, which produces the maximum possible quantum correlation.  As $r$ increases, the post-selected correlation functions approach those of a PR box \cite{popescu94}, which takes on the form of a square wave. Choosing $\beta = \pi / 4$, $\beta^\prime = 3 \pi / 4$ and $r \gtrsim 1$mm, the violation of the CHSH inequality exceeds the Tsirelson bound. 

\begin{figure}
\includegraphics[width=7cm]{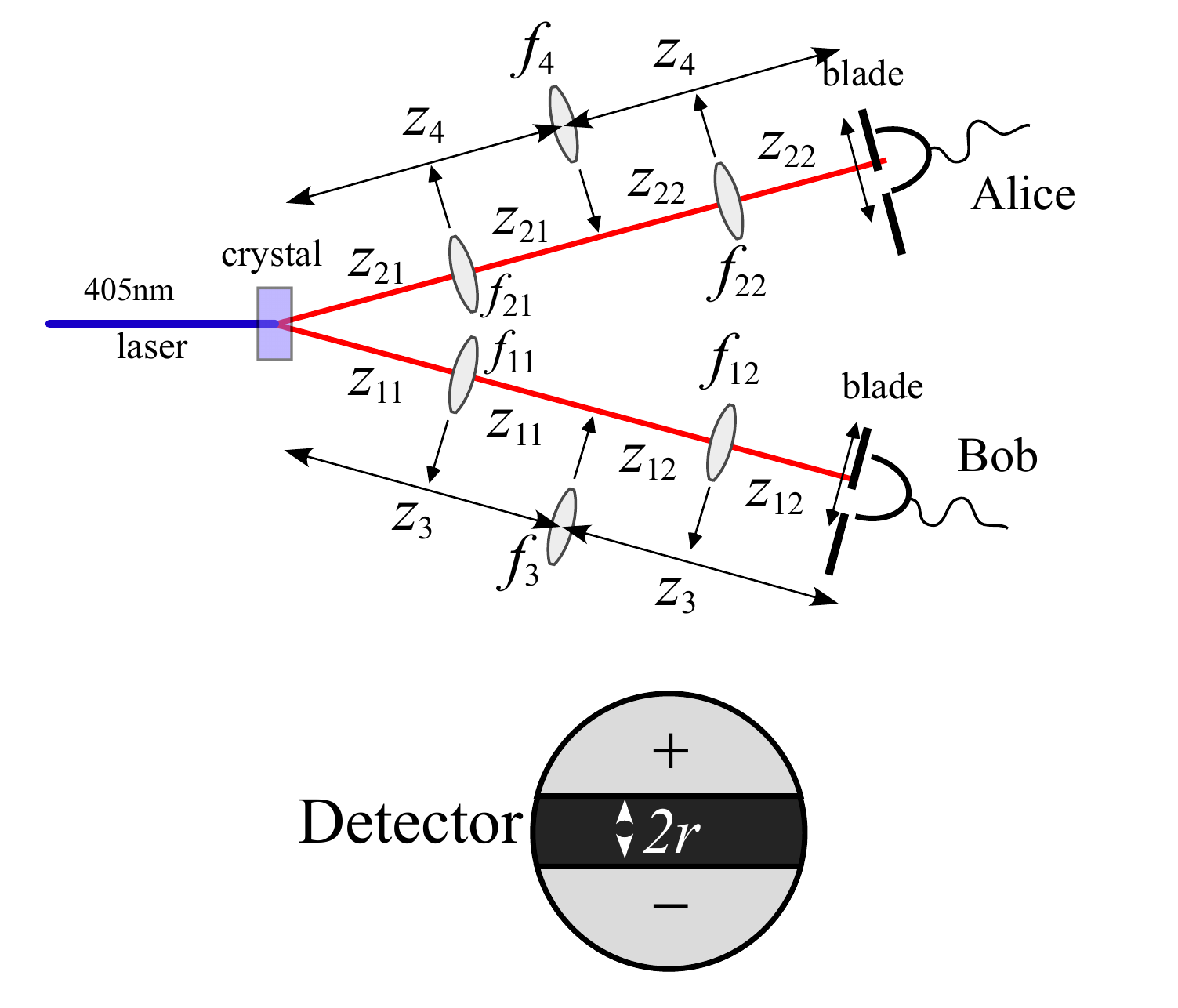}
 \caption{\label{fig:setup} (Color online) Experimental setup and measurement scheme.}
 \end{figure}

Figure \ref{fig:setup} shows the experimental setup.   A 50mW diode laser ($\lambda\sim405$ nm) is first sent through a $\times 3$ beam expander and is then used to pump a 5 mm long BBO crystal cut for type I phase matching, producing spatially entangled photon pairs by SPDC.  The down-converted signal and idler photons are detected by single photon counting modules equipped with 10 nm FWHM interference filters centered around 810 nm. The state describing the spatial degrees of freedom of the photons is approximately described by Eq. \eqref{eq:state}. The phase space rotations \eqref{eq:rotations} are realized via the fractional Fourier transform (FRFT), which can be implemented with optical lens systems \cite{lohmann93,tasca08,tasca09}: a lens with focal length $f$ placed symmetrically at distance $z_\theta=2f\sin^2(\theta/2)$ from both the input and output plane realizes a $\theta$-order FRFT, which corresponds to a rotation of angle $\theta$ in the phase space composed of dimensionless position and wave vector coordinates.  The lenses used in the optical FRFT systems were mounted on detachable mirror mounts so that they could be toggled in and out of the paths of the down-converted photons, while preserving the alignment. Rotations of approximately $\alpha=\pi$ and $\alpha^\prime=\pi/2$ were implemented on the signal photon (Alice's side) and $\beta=5\pi/4$ and $\beta^\prime=3\pi/4$ on the idler photon (Bob's side).  Using the additivity property of the FRFT \cite{ozaktas01}, the $\alpha\approx \pi$ and $\beta \approx 5\pi/4$ FRFTs were composed of two FRFTs, $\alpha=\alpha_{1}+\alpha_{2}$ and $\beta=\beta_{1}+\beta_{2}$, as shown in figure \ref{fig:setup}.
The focal lengths and $z$-distances used are shown in table \ref{tab:1}. 
  \begin{table}
 \caption{\label{tab:1}Focal lengths $f$ and distances $z$ for optical FRFTs of order $\theta$.}
 \begin{ruledtabular}
 \begin{tabular}{ccc}
angle & $l$ & $z$ \\
\hline
$\alpha_{1} = \frac{\pi}{2}$& $f_{11}$=25.0cm & $z_{11}=$25.0cm  \\
 $\alpha_{2} = \frac{29\pi}{50}$ & $f_{12}=$20.0cm & $z_{12}=$25.0cm \\
$\beta_{1}= \frac{\pi}{2}$ & $f_{21}$=25.0cm & $z_{21}=$25.0cm \\
$ \beta_{2}=\frac{3\pi}{4}$ & $f_{22}=$15.0cm & $z_{22}=$25.6cm \\
$\alpha^\prime=\frac{\pi}{2}$ & $f_{3}=50.0$cm & $z_{3}=50.0$cm \\
$\beta^\prime=\frac{37 \pi}{50}$ & $f_{4}=30.0$cm &  $z_{4}=50.6$cm \\
\end{tabular}
 \end{ruledtabular}
 \end{table}
  \begin{figure}
\includegraphics[width=5.5cm]{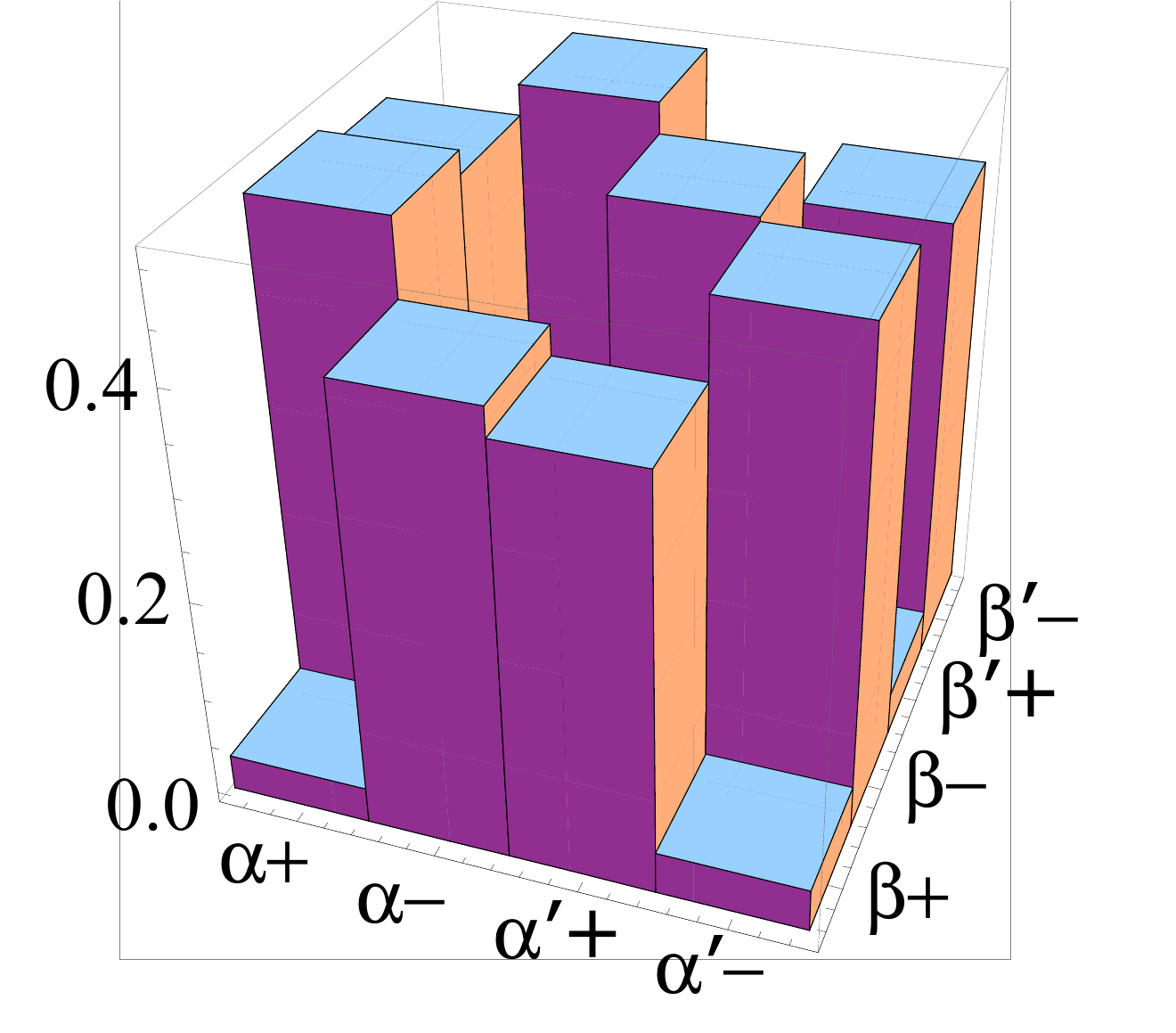}
 \caption{\label{fig:results} (Color online) Probabilities for each measurement combination for $r=0.5$mm.}
 \end{figure}
\par
We measured the coincidence counts for the four combinations of rotations for three values of $r$.  The joint probabilities were calculated by normalizing the coincidence counts. Figure \ref{fig:results} shows the measured probabilities for $r=0.5$mm.  The marginal probability distributions for Alice or Bob can be obtained by summing up the rows or columns and dividing by two. For example, the figure shows that for Alice's measurement $\alpha$ the probabilities to obtain result ``$+$" and ``$-$" are 0.503 and 0.497, respectively.  All marginal probabilities are approximately equal to $1/2$, thus fulfilling the no-signalling condition. Without post-selection ($r=0$), no violation of the CHSH inequality is  
obtained.
\par
With $r=0.2$mm, it was already possible to surpass Tsirelson's bound.  By increasing $r$, a larger violation was obtained.  Table \ref{tab:2} summarizes the measurement results for $r=0.1$mm, 0.2mm and 0.5mm.  The largest violation obtained was about $\mathcal{S}=3.42$.   $H_{ave}$ expresses the average percentage of collected pairs in each configuration.     
\par
As an application of our results we note that the experiment can be used to implement a non-local AND gate.  By identifying Alice and Bob's rotations as bit values--say--$\alpha,\beta \rightarrow 0$ and $\alpha^\prime,\beta^\prime \rightarrow 1$--the post-selected measurement results implement an AND gate, where the sum modulo 2 of the outcomes of Alice and Bob represent the output bit and we identify ``$+$" (``$-$") measurement as the logical vaule $0$ ($1$).  Negative values for the correlation function are obtained only when both Alice and Bob choose the rotation angles $\alpha^{\prime} $ and $\beta^{\prime} $, respectively. The success probability of the post-selected AND gate is given by $P_{\mathrm{AND}}=\frac{1}{4}[P_{\alpha \beta}(+,+)+P_{\alpha \beta}(-,-) + P_{\alpha^{\prime} \beta}(+,+) + P_{\alpha^{\prime} \beta}(-,-) + P_{\alpha \beta^{\prime}}(+,+)+P_{\alpha \beta^{\prime}}(-,-) + P_{\alpha^{\prime} \beta^{\prime}}(+,-)+P_{\alpha^{\prime} \beta^{\prime}}(-,+)]$, and is also shown in table \ref{tab:2}. The highest success probability obtained is around $0.93$.  
  \begin{table}
 \caption{\label{tab:2} Post-selected correlation functions for different values of $r$, along with value of Bell parameter $\mathcal{S}$ and success probability of a non-local AND gate.}
 \begin{ruledtabular}
 \begin{tabular}{cccccccc}
r  & $H_{ave}$ & $E(\alpha,\beta)$ & $E(\alpha^\prime,\beta)$ & $E(\alpha,\beta^\prime)$ & $E(\alpha^\prime,\beta^\prime)$ & $\mathcal{S}$ & $P_{\mathrm{AND}}$ \\
\hline
0.1 mm   & 84\% & 0.717 & 0.862 & 0.517  & -0.743 & 2.84 & 0.85 \\
0.2 mm  & 74\% & 0.721 & 0.899 & 0.580 & -0.802 & 3.00 & 0.88 \\
0.5  mm & 41\% & 0.755 & 0.975 & 0.828 & -0.860 & 3.42 & 0.93 \\
\end{tabular}
 \end{ruledtabular}
 \end{table}
\par
Through post-selection of a two-mode Gaussian state we were able to simulate non-local PR correlations.
Post-selection allowed us not only to violate CHSH inequality, but also to surpass the Tsirelson's bound. The greater the violation of the CHSH inequality, the higher the fidelity of the implementation of the PR box. We achieve a maximum fidelity of 93\%. This fidelity is higher than the upper bound of 90.8\% \cite{brassard06}, below which the communication complexity is non-trivial and above which it is trivial. We have used the transverse degrees of freedom of twin photons, produced in parametric down-conversion, in a spatially binned configuration. However, these concepts can be easily extended to other continuous variable systems. We expect that these results will allow for the further experimental  
investigation of PR correlations and a wide variety of fundamental aspects of physics such as non-local  computation \cite{linden07}, the no-signaling theorem, distillation of correlations and communication complexity \cite{foster09,brunner09}. Our results are also a clear demostration of the importance played by any detection loophole in a legitimate test of Bell's inequality.
\par

D. S. Tasca thanks D. Cavalcanti for stimulating discussions.  Financial support was provided by Brazilian agencies CNPq, PRONEX,
CAPES, FAPERJ, FUJB and the Institutos Nacionais de Ci\^encia e Tecnologia de Informa\c{c}\~ao Qu\^antica (INCT).



\end{document}